\documentclass[12pt,rotating]{elsart}
\usepackage{rotating}

\usepackage{cites,setspace,url}
\usepackage{epsfig}
\usepackage{doublespace}

\bibnum{k}


\begin{document}

\citenum{}

\begin{frontmatter}
\title{Physics process of cosmogenics $^9$Li and $^8$He production on muons 
interactions with carbon target in liquid scintillator}

\author{Karim Zbiri$^{*~1,2}$}
\corauth{ Tel.: +1 215 895 2990.\\  E-mail address: zbiri@physics.drexel.edu\\
(K. Zbiri). }

\maketitle

\address{$^1$Department of physics, Drexel University, 3141 Chestnut St, 
Philadelphia, PA 19104, USA}
\address{$^2$Laboratoire Subatech, UMR6457: IN2P3/CNRS-universite-Ecole des 
Mines de Nantes, 4 rue A. Kastler, 44307 Nantes Cedex 03, France}

%\maketitle

\begin{abstract}
Simulations were performed with Geant4 to study the cosmogenics $^9$Li and 
$^8$He production by the interactions of the muons and their secondary
shower particles in the liquid scintillator. The photonuclear reactions seem 
dominate their production. Their energy dependence were deduced. The results
of the simulations are compared with available data.
\\

\end{abstract}

\end{frontmatter}

\section{INTRODUCTION}

In reactor neutrino experiments the standard detection reaction is
the inverse beta decay : \[
\overline{\nu_{e}}+\, p\,\rightarrow n\,+\, e^{+}\]

The positron makes a prompt signal with energy $\sim$ 1 to 8 MeV and
the neutron produce a delayed signal after capture in gadolinium with
energy $\sim$ 8 MeV. The detection signal can be mimicked by a radioactive
isotopes (like $^8$He, $^9$Li and $^{11}$Li) which decay by $\beta$-neutron
cascade. This kind of isotopes is produced by the cosmic muons and their 
shower particles interactions  
with the carbon nuclei which is the main component
of the liquid scintillator. Because these isotopes are produced by
cosmic muons interactions they are called cosmogenics. The main problem
with the cosmogenics is due to their long life time (178 ms for $^9$Li
and 119 ms for $^8$He) which make them untagged by the veto system. Therefore
the simulation to evaluate their rate for each reactor neutrino experiment
becomes crucial. The knowledge of the physics processes behind cosmogenics
production is of first interest in order to perform their rate simulation.

\section{$^9$Li and $^8$He YIELD}

The importance of the
electromagnetic and hadronic cascades in the neutrons production in the liquid 
scintillator
are discussed in different works \cite{Wang01,Araujo05}. In this
paper I investigate on their importance on the cosmogenics $^9$Li and $^8$He 
production.

\begin{figure}[!h]
\begin{center}\includegraphics[%
  scale=0.8]{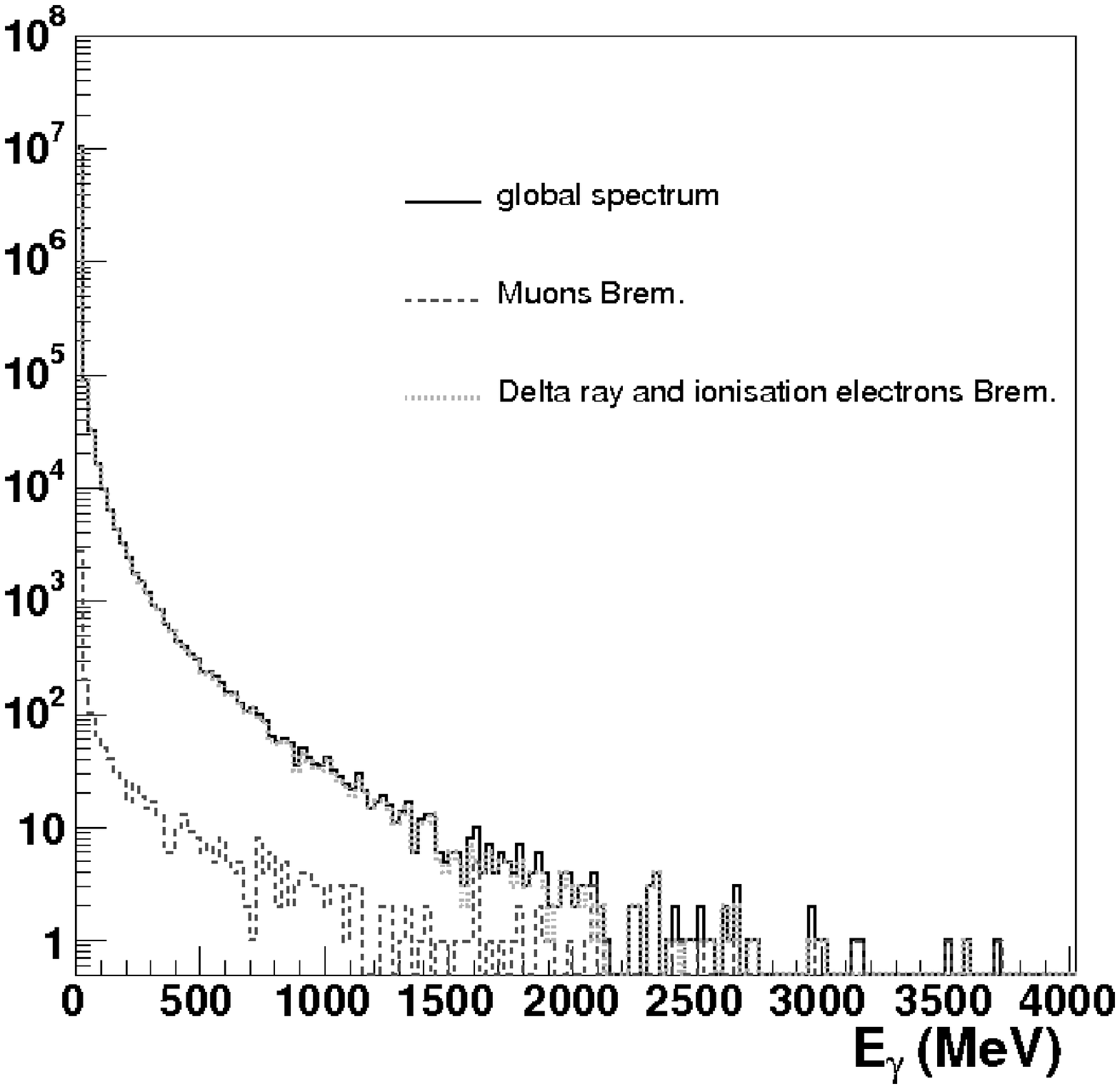}\end{center}
\caption{\label{cap:Gamma-spectra-produced}Gamma spectra produced by 
Bremsstrahlung process from interactions of 50000 muons with energy of 10 GeV 
through 10 meters on the rock.}
\end{figure}

One physics process which can contribute to the cosmogenics production in the 
liquid 
scintillator is the real photonuclear reaction. The gammas source is the muons
and their shower particles Bremsstrahlung. As shown in 
Fig. \ref{cap:Gamma-spectra-produced}, where are
plotted different gammas spectra, one from muons Bremsstrahlung and
the other from delta ray and ionization electrons Bremsstrahlung,
the gammas are mainly produced by muons shower particles Bremsstrahlung.
The gammas will be produced on the surrounding rock and propagate
inside the detector, the use of shielding will minimize their impact,
also the gammas can be created inside the detector but the production
of gammas in this case is not as important as the one on the rock.

\begin{figure}[!h]
\begin{center}\includegraphics[%
  scale=0.6]{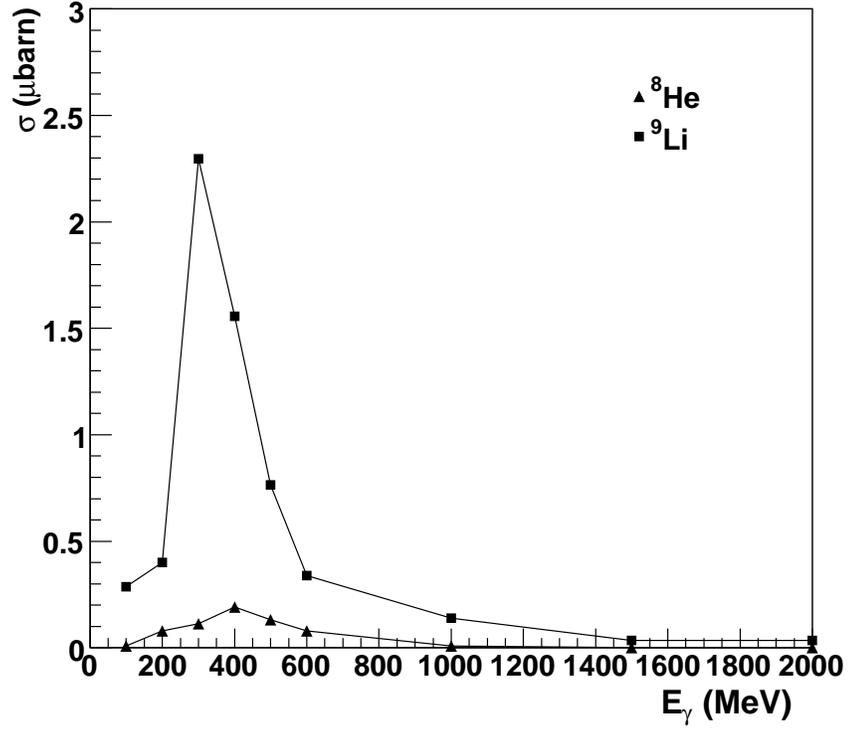}\end{center}
\caption{\label{cap:Cross-section-LiHe}Cross section of $^9$Li and 
$^8$He production from interactions of $10^{7}$ gammas through 1 meter 
on the $^{12}C$.}
\end{figure}

\begin{figure}[!h]
\begin{center}\includegraphics[%
  scale=0.6]{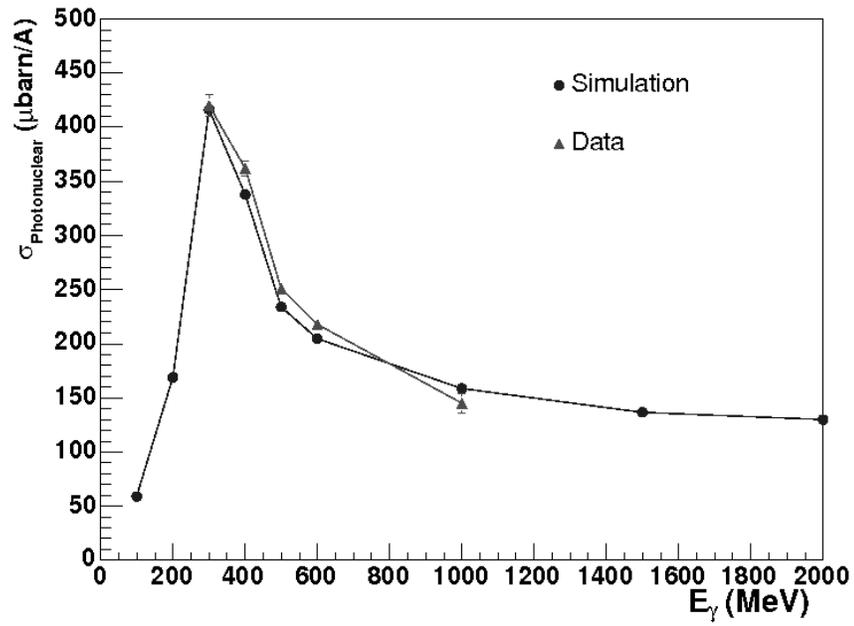}\end{center}
\caption{\label{cap:Photonuclear-total-cross}Photonuclear total cross section
normalized by the mass number A as a function of gamma energy, obtained
from interaction of $10^{7}$gamma on 1 meter of $^{12}C$.}
\end{figure}

To evaluate the yield of $^9$Li and $^8$He, I performed a simulation
with Geant4 Monte Carlo code (release 7.1) \cite{Geant4}. The geometry
used is a simple cube made of $^{12}C$ in which I shot a beam of $10^7$ 
monochromatic energy gammas. To cover different interaction regions, the 
gammas energies were chosen like following:
20 MeV, 30 MeV, 50 MeV, 80 MeV, from 100 MeV to 1000 MeV by step of 
100 MeV and from 1 GeV to 4 GeV by step of 0.5 GeV. I obtained $^9$Li and 
$^8$He production only in the region corresponding to the interval of energy
going from 100 MeV to 2 GeV.
The Figure \ref{cap:Cross-section-LiHe}
shows the cross section obtained from the simulation as a function of the gamma
energy. The variation of the energy goes from 100 MeV to 2 GeV, this energy
range cover three different interaction regions: $\Delta$ region
(from the pion threshold to 450 MeV), Roper resonance region (from
450 MeV to 1.2 GeV) and Reggeon-Pomeron region ($>$ 1.2 GeV). Below 3
GeV, to generate the final state for gamma-nuclear inelastic scattering,
the real photonuclear interaction, Geant4 uses a chiral-invariant
phase space model \cite{Deg01}. In Geant4 the photonuclear cross
sections are parametrized from several measurements performed for
many nuclei and cover a large range of gamma energy \cite{Kossov02}. 
Figure \ref{cap:Photonuclear-total-cross}
shows the comparison between the photonuclear cross sections obtained
from simulating the interactions of $10^{7}$ gammas on 1 meter of $^{12}C$
and one set of available data \cite{Bian96}, we can observe the good
agreement between the simulation and data, more comparisons were made in the 
Ref. \cite{Kossov02}.

As we can observe from Fig. \ref{cap:Cross-section-LiHe}, $^9$Li is more 
likely produced than $^8$He
and the cross section of the two isotopes increase with the gamma energy 
to reach
its maximum around 2.3 $\mu barn$ for a gamma energy about 300-400 MeV 
and decrease with
increasing energy. In fact the evolution of the cosmogenics cross section 
follows the
total photonuclear cross section behavior and the maximum of cosmogenics
cross section corresponds to the maximum of total real photonuclear reaction
cross section as shown in Fig. \ref{cap:Photonuclear-total-cross}.

Only one experiment was dedicated to investigate the cosmogenics production
in muons beam interaction with scintillator target at SPS \cite{Hag00}.
During this experiment no distinction between $^8$He and $^9$Li was possible
due to their neighboring half-lives. The measured cross section for
$^9$Li and $^8$He with 190 GeV muons beam is $\sigma(^{9}Li+^{8}He)=(2.12\pm0.35)$$\mu barn$.

\section{ENERGY DEPENDENCE OF THE $^9$Li and $^8$He YIELD}

At SPS experiment the energy dependence of the total cross-section
was evaluated to be $\sigma_{tot}(E_{\mu})\propto E_{\mu}^{\alpha}$,
$\alpha$ varies from 0.50 to 0.93 with a weighed mean value $<$$\alpha$$>$
= 0.73 $\pm$ 0.10. In order to extract the value of $\alpha$, the
ratio of the total cross section at 100 and 190 GeV was performed.
Nevertheless, the total cross section for $^9$Li and $^8$He production
was obtained only for 190 GeV muon energy, thus these two last isotopes
were not taken into account to determine the $\alpha$ parameter. This energy
dependence of the isotopes production is well known as described in the 
Ref. \cite{Khal95}.\\
In
the current section I describe how I obtained the $\alpha$ value, corresponding
to the cosmogenics $^{8}He$ and $^{9}Li$ production, via appropriate
simulation. I used like a target a cylindrical bloc of scintillator made from dodecane
(e.g. $C_{12}H_{26}$), the cylinder has radius of 1.7 meter and high of 
3.5 meters. 
I put the target in the center of 
a bloc of the rock ($SiO_2$ cube of 6 meters size) in which I shot 
a $10^6$ monochromatic energy muons. Since the energy dependence of the 
isotopes production is valid
for a wide range of muons energy \cite{Khal95}, it can be checked with any different energies chosen within this range. Therefore, I performed
two simulations one with muon energy of 100 GeV and the second of
500 GeV. All the electromagnetic and hadronic processes were allowed
for the muons and their secondary shower particles. The history of 
each produced isotope was traced back,
but all the $^9$Li and $^8$He formed inside the target scintillator were only
produced from the photonuclear reactions, the other processes like 
direct muon-carbon interaction, electron-carbon or neutron-carbon interaction
didn't produce any $^9$Li or $^8$He. Table \ref{cap:Yield-LiHe-munrj}
shows the obtained yield (the number of isotopes produced in each run) for 
each isotope produced in the target (the bloc scintillator) and the
value of the $\alpha$ parameter at 100 and 500 GeV. To obtain the value of 
$\alpha$ I used the ratio of the simulated yield at 100 and 500 GeV, in the 
same way like in SPS experiment:

\[
%\begin{equation}
\alpha\,=\,\frac{ln(yield(500\, GeV)/yield(100\, GeV))}{ln(500\, GeV/100\, GeV)}
%\end{equation}
\]

The values of $\alpha$ are in good agreement, within the error bars,
with experimental ones \cite{Hag00,Khal95}.

\section{CONCLUSION}
Interacting cosmic muons initiate electromagnetic and
hadronic cascades which dominate neutrons production \cite{Araujo05}.
In the case of the cosmogenics $^{8}He$ and $^{9}Li$, the simulations 
with Geant4 shows that the real photonuclear reaction seems to be the 
main physics process in their production. 
The gammas are mainly produced by Bremsstrahlung
of muons shower particles on the rock. After their creation the gammas
propagate inside the detection volume where they interact with the carbon
target to produce radioactive isotopes and secondary particles. The
simulated values for $^9$Li and $^8$He cross sections are in agreement
with SPS experiment. Also, the
simulated energy dependence agree with those obtained from different experiments \cite{Hag00,Khal95}.

%\newpage
\section*{Acknowledgments}

I would like to thank Prof. Jacques Martino for his several helpful advices and discussions through this work.

\newpage

\newpage

\newpage

\newpage

\begin{table}[!h]
\begin{center}\begin{tabular}{|c|c|c|c|}
\hline 
Isotopes&
yield at 100 GeV&
yield at 500 GeV&
$\alpha$ parameter\tabularnewline
\hline
$^9$Li&
7&
38&
1.06\tabularnewline
\hline 
$^8$He&
2&
7&
0.78\tabularnewline
\hline
\end{tabular}\end{center}

\caption{\label{cap:Yield-LiHe-munrj}The yield (number of isotopes produced in 
each run) of $^9$Li and $^8$He produced in the target (the bloc scintillator) 
and the $\alpha$ value obtained for the different simulated energies.}
\end{table}

\end{document}